\documentclass[aps,longbibliography,notitlepage,floatfix,11pt,tightenlines,nofootinbib]{revtex4-1}

\usepackage{bm,amsmath,amssymb,graphicx,stmaryrd,subfigure}
\usepackage[abs]{overpic}
\newcommand{\uvc}[1]{\bm{\mathrm{\hat #1}}} 

\newcommand{\bX}{{\bf X}}
\newcommand{\bP}{{\bf P}}
\newcommand{\bR}{{\bf R}}
\newcommand{\bC}{{\bf C}}
\newcommand{\bO}{{\bf O}}

\newcommand{\RL}{{\begin{smallmatrix}R\\L\end{smallmatrix}}}

\begin{document}

\title{Jump conditions for strings and sheets from an action principle}
\author{J A Hanna}
\email{hannaj@vt.edu}
\affiliation{
Department of Biomedical Engineering and Mechanics,\\
Department of Physics,\\
 Virginia Polytechnic Institute and State University;\\
\mbox{Norris Hall (MC 0219), 495 Old Turner Street, Blacksburg, VA 24061, U.S.A.}}

\date{\today}

\begin{abstract}
I present conditions for compatibility of velocities, conservation of mass, and balance of momentum and energy across moving discontinuities in inextensible strings and sheets of uniform mass density.  The balances are derived from an action with a time-dependent, non-material boundary, and reduce to matching of material boundary conditions if the discontinuity is stationary with respect to the body.  I first consider a point discontinuity in a string and a line discontinuity in a sheet, in the context of classical inertial motion in three Euclidean dimensions.  I briefly comment on line discontinuities terminating in point discontinuities in a sheet, discontinuous line discontinuities in a sheet, and an approach to dynamic fracture that treats a crack tip in a sheet as a time-dependent boundary point.  I provide two examples of general solutions for conservative sheet motions near a line discontinuity.  The approach also enables treatment of actions depending on higher derivatives of position; I thus derive balances for an \emph{elastica} which are applicable to moving contact problems.  
\end{abstract}

\maketitle

The dynamics of strings and sheets offer many surprises \cite{HannaKing11, Cambou12, CoteronVimeo, JuddVimeo, MouldSiphon2, Biggins14, Virga14, Rennie72, Calkin89, Schagerl97, Tomaszewski06, HammGeminard10, Taneda68, Bejan82, Guven13skirts}.  Their motion unregularized by any resistance to bending, these perfectly flexible, yet inextensible bodies may develop persistent kinks and other discontinuities in their shapes.  Embedded in three-dimensional space, they come into partial contact with steric, frictional, and adhesive obstacles, and thereby experience discontinuous external applied forces.  Thus, the mechanics of such discontinuities should be studied. 

\begin{figure}[here]
	\includegraphics[width=6in]{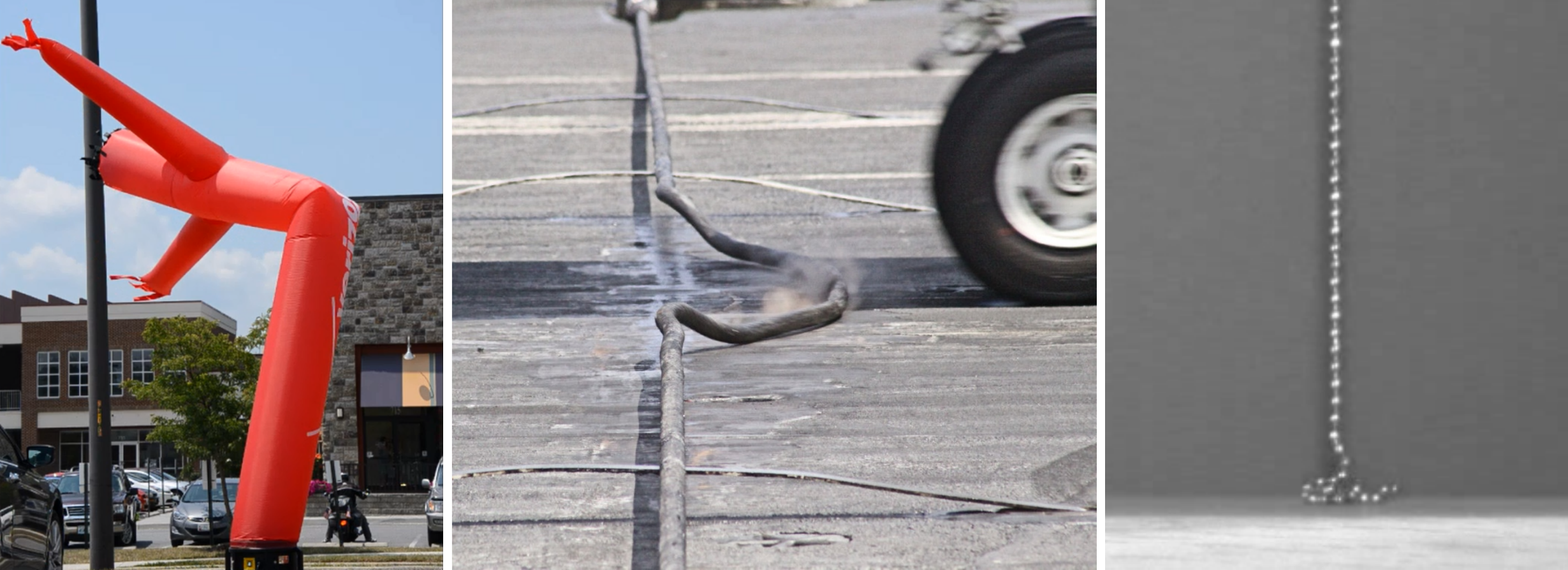}
\caption{Real systems with discontinuous, or approximately discontinuous, moving features.  Left: A kink moving upward in a tubular  Airdancer$^\mathrm{\textregistered}$ \cite{Airdancer} membrane (still from a film courtesy R B Warner).  Center: A propagating impact in an aircraft arresting cable \cite{ArrestingGearWikipedia}.  Right: A chain falling onto a table (still from a film courtesy D Aliaj and R B Warner).  The table, or the pile of chain on the table, acts as a positive supply of stress (momentum) and a negative supply of power (energy).}
\label{realjumps}
\end{figure}

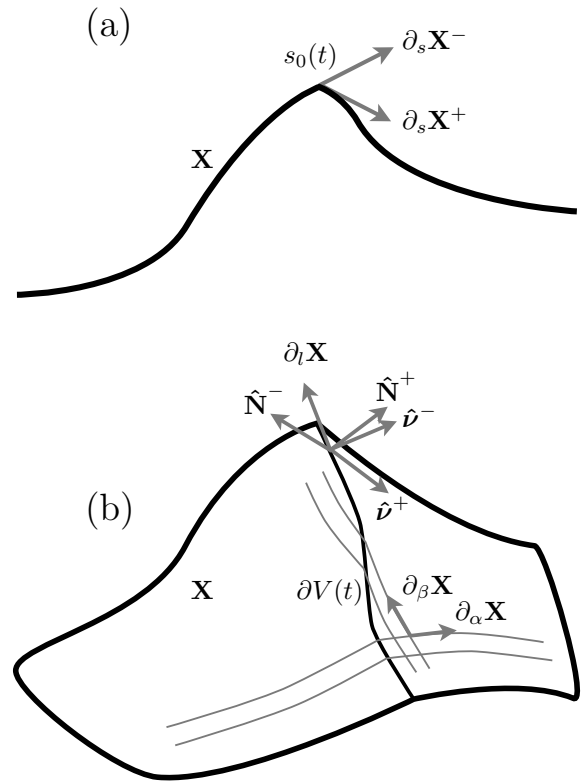
\begin{figure}[here]
\subfigure{
	\begin{overpic}[width=3in]{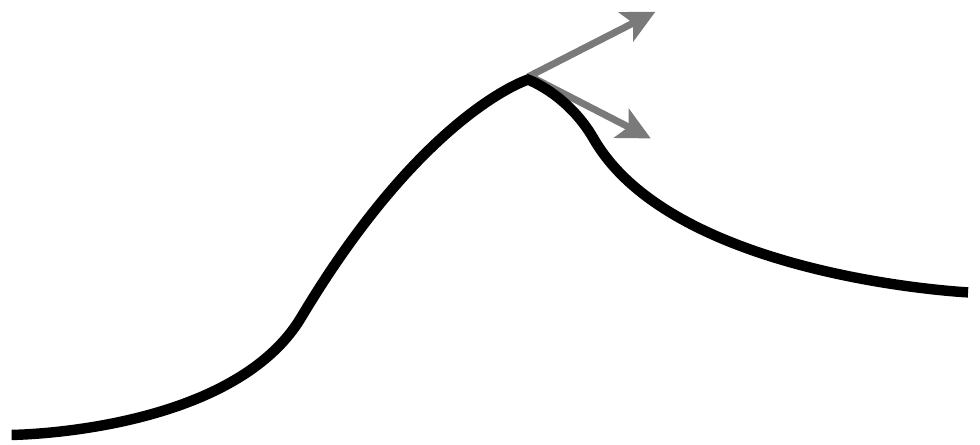}\label{curve}
	\put(25,100){\Large{\subref{curve}}}
	\put(65,50){$\bX$}
	\put(145,95){$\partial_s\bX^-$}
	\put(145,65){$\partial_s\bX^+$}
	\put(100,90){$s_0(t)$}
	\end{overpic}
	}
\\\vspace{.25in}
\subfigure{
	\begin{overpic}[width=3in]{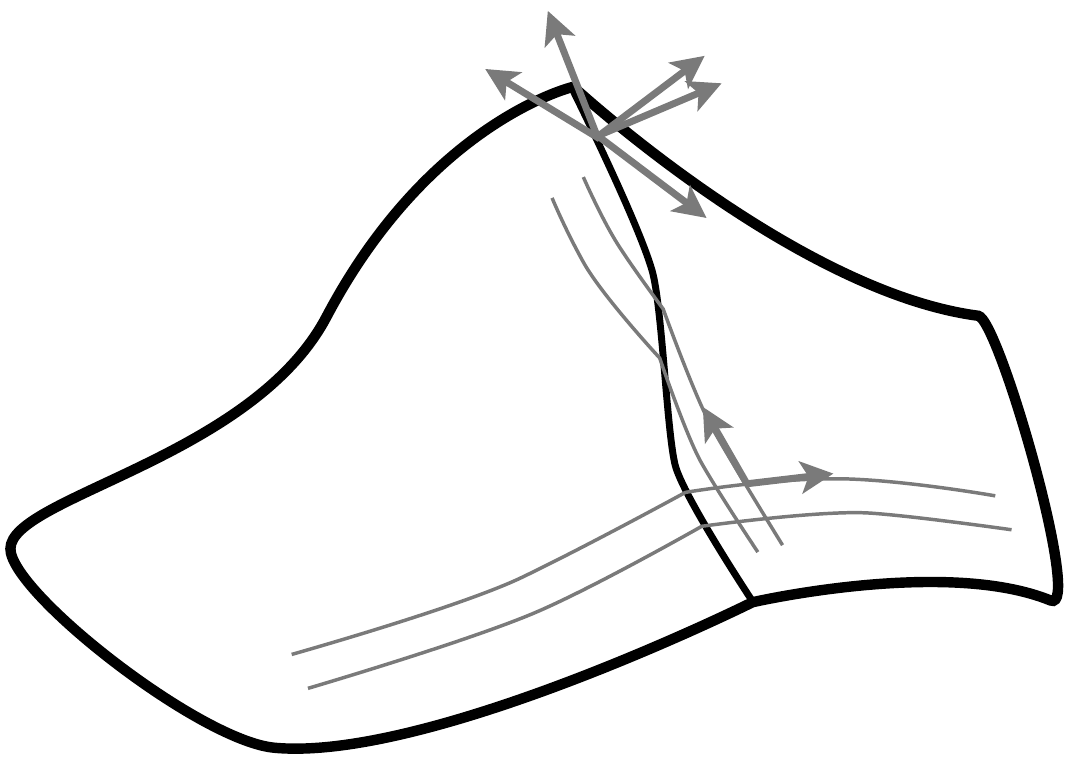}\label{surface}
	\put(25,100){\Large{\subref{surface}}}
	\put(65,70){$\bX$}
	\put(165,62){$\partial_\alpha\bX$}
	\put(145,72){$\partial_\beta\bX$}
	\put(145,135){$\uvc{\nu}^-$}
	\put(135,100){$\uvc{\nu}^+$}
	\put(85,140){$\uvc{N}^-$}
	\put(135,145){$\uvc{N}^+$}
	\put(100,160){$\partial_l \bX$}
	\put(105,70){$\partial V(t)$}
	\end{overpic}
	}
\caption{Bodies $\bX$ with discontinuities, here drawn as first order in derivatives of position.  \subref{curve} Tangent vectors at a point discontinuity in a string.  \subref{surface} Tangent vectors and Darboux frames at a line discontinuity in a sheet, and coordinate lines and tangents in the bulk.  Details and definitions in text.}
\label{jumps}
\end{figure}

In this paper, I consider moving discontinuities in one- and two-dimensional flexible bodies (see Figures \ref{realjumps} and \ref{jumps}).  The bodies are modeled as inextensible curves and surfaces.  Physical examples of such discontinuities include peeling fronts of adhesive tapes and coatings \cite{Ericksen98, BurridgeKeller78, Cortet13, HureAudoly13}, lift-off points of chains and ropes moving around pulleys or table edges \cite{Rennie72, PratoGleiser82, Calkin89, Cambou12}, pick-up points of chains from piles or rigid surfaces \cite{Cayley1857, HannaKing11, Biggins14, Virga14, Virga15}, propagating impacts in cables and membranes \cite{Ringleb57, Cristescu64, BeattyHaddow85, Yokota01, Tomaszewski06, HannaSantangelo12, VandenbergheVillermaux13, KanninenFlorence67, Farrar84, Haddow92, AlbrechtRavi-Chandar14}, geometrically complex propagating kinks in a windblown flag or the tubular body and arms of an Airdancer$^\mathrm{\textregistered}$ \cite{Airdancer}, brittle cracks, tears, and cuts in sheet materials \cite{BurridgeKeller78, Roman13, VandenbergheVillermaux13}, and groove structures in impressed bladders.  It also seems likely that kinks may form in the transverse waves resulting from hairpin turn maneuvers of towed cables \cite{IversMudie73, Sanders82, Matuk83}.  In some of these examples, the discontinuity is an idealization that allows one to ignore a regularizing length scale, such as the body thickness, and with it higher-order terms in equations of motion, such as bending forces.  When considering an elastic sheet, it may be easier to treat membrane equations and a jump condition than to explicitly examine an internal bending boundary layer and perform asymptotic matching.  Thus, jump conditions may be an effective tool in studies of thin elastic bodies.  For dynamic partial contact problems involving continua, the correct jump conditions are essential; one cannot simply apply static boundary conditions.

The first conditions that must hold across a discontinuity are compatibility of velocities and some body derivatives of position.  A treatment of surfaces generates the equivalents of Hadamard and Darmois compatibility, which differ from the classic cases because both the discontinuity and its embedding surface are of nontrivial codimension with respect to another embedding space, $\mathbb{E}^3$.

\clearpage

Next I consider mass conservation, a topic which is made considerably easier by the presumption that the bodies are inextensible and isometric to a sufficiently smooth configuration.  This justifies the use of global material coordinates on the body which facilitate calculations.  

Then I turn to balances of momentum and energy.  The balances are derived from an action with a time-dependent, non-material boundary, rather than from a set of conservation laws in weak form as would be traditional in continuum mechanics.  For this reason, the formulation shares some conceptual ground with variable-mass problems \cite{McIver73}, such as axially moving belts between supports \cite{WickertMote88, LeeMote97-1}, yarn or cables deployed off of spools or onto the seafloor \cite{Mack58, Zajac57, Padfield58, MankalaAgrawal05, Krupa06}, pipes conveying fluid \cite{PaidoussisLi93}, or other situations in which one or more boundaries act as sources or sinks of material. 

This paper is restricted to metrically constrained inertial motions in $\mathbb{E}^3$, but the method can be generalized to other systems involving elasticity or plasticity of the body. 
Aside from the economy of assumptions inherent to a variational principle, there seem to be conceptual advantages to viewing a discontinuity as a moving boundary, rather than an internal ``wave''. 
There is the possibility of treating fracture, as well as combined line and point defects, by the approach suggested in Section \ref{crack}.   The ability to consider boundary conditions of action functionals of arbitrarily high derivatives of position is exploited in Section \ref{elastica} to derive momentum and energy balances for a discontinuous \emph{elastica}.  Finally, the present treatment adds a new perspective on the existence of energy functionals for some axial motions of thin bodies between supports \cite{Renshaw98}.

The prior work of McIver \cite{McIver73} should be mentioned, in which an action principle was developed for an open system akin to the time-dependent volumes considered in the present work.  The present treatment differs primarily in its focus on boundary conditions and their application to discontinuities, and in its use of global material body coordinate descriptions of the moving boundaries.

The variations in the current procedure involve only the material position vector and the time; it is likely that these could be viewed as a single four-dimensional material position vector.  This is in contrast to variations of the position of geometric quantities, such as the location of a boundary or other defect.  Examples of the latter may be found in recent treatments of static adhesion \cite{Deserno07, MajidiAdams09, Majidi12, HureAudoly13}, and in approaches based on the concept of configurational balances \cite{Gurtin00, KienzlerHerrmann00, Maugin11}.  The current procedure is more direct, not requiring any additional compatibility conditions on the variations.  Perhaps more importantly, it does not rely on any principles beyond the established action of classical mechanics that applies to the material composing the body; there are no new postulated laws for geometric objects.  However, the approach is limited to defects that can be described as boundaries, excluding non-Riemannian objects such as dislocations.

A variational approach to momentum and energy equations in imperfect elastic continua was pioneered by Eshelby \cite{Eshelby51, Eshelby75}.  His original focus was on non-Riemannian defects such as dislocations, which any variational integral contour must necessarily surround but not pass through.  This approach of using an encircling contour was later applied to cracks \cite{AtkinsonEshelby68}, mirroring other work in fracture mechanics \cite{Rice68, Freund90, Ravi-Chandar04}.  I will suggest in Section \ref{crack} that a crack tip may be approached more directly as the time-dependent boundary of a boundary of an action integral.

The illustrations in Figure \ref{jumps} show discontinuities in the tangent vectors of bodies, which are the relevant discontinuities of the simple actions considered in the primary sections of this paper, \ref{string} and \ref{sheet}.  However, it should be emphasized that the present method is not restricted to first order discontinuities in position.  
One expects higher order discontinuities to appear in many problems featuring bending forces, such as dynamic rolling or sliding of elastomeric or \emph{elastica}-like materials \cite{Schallamach71, StolteBenson93, Raux10}, unfurling tape springs with localized eversions \cite{SeffenPellegrino99}, space-fixed transverse loads on spinning disks or moving tapes \cite{BensonBogy78, RenshawMote95, Albrecht77}, and the impact of liquid or solid rods and sheets with surfaces \cite{MahadevanKeller95, MahadevanKeller96, Ribe03, Ribe04, Habibi07}.  Treatments of extended defects or surgery of two space-times in general relativity \cite{BonnorVickers81, Vilenkin85, Blau87} involve discontinuities in higher order derivatives of a metric.  In the last case, there is no analogue of our second, Euclidean embedding space, or of material coordinates, but it is still fruitful to work in a coordinate system that smoothly spans the discontinuity \cite{Blau87}, as in the present paper.  

The concepts and approach of this paper are first introduced in the simple context of a string in Section \ref{string}, before being applied to sheets in Section \ref{sheet}.  I recover known results for a point discontinuity in a string, and present new results for a line discontinuity in a sheet.  These results coincide with, or are analogous to, known results derived from conservation laws in weak form for strings and space-filling bodies of trivial codimension.    I use as comparative references a paper of O'Reilly and Varadi \cite{OReillyVaradi99} and a text by Gurtin, Fried, and Anand \cite{GurtinFriedAnand10}.  
  In Section \ref{crack}, I briefly comment on line discontinuities terminating in point discontinuities in a sheet, discontinuous line discontinuities in a sheet, and the treatment of a crack tip in a sheet as a time-dependent boundary point.  In Section \ref{examples}, I provide two solutions of conservative sheet motions near a line discontinuity.  In Section \ref{elastica}, I derive balances for an \emph{elastica}.\\


\section{A point discontinuity in a string}\label{string}

In this section I will obtain, by an alternate method, a restricted form of some results derived by O'Reilly and Varadi \cite{OReillyVaradi99}.  Their work, based on that of Green and Naghdi \cite{GreenNaghdi79-1, GreenNaghdi78}, was also further developed in a later paper \cite{OReillyVaradi03}.  

A possible configuration of a one-dimensional body $\bX(s,t)$ is shown in Figure \ref{curve}, where $\bX$ is a position vector in $\mathbb{E}^3$ parameterized by arc length $s$ and time $t$.  The body has a discontinuity, located at the time-dependent coordinate $s_0(t)$.  It is simplest, though not necessary, to consider that the discontinuity moves through a global material coordinate system covering the entire body, so that the description of $s_0(t)$ is common to both sides of the body.  In the figure, the discontinuity is depicted as being in the first derivative of the position vector, leading to a jump in the unit tangent vector $\partial_s\bX$.  Higher order discontinuities in position may be considered in the same manner.  

A jump will be written using double square brackets, 
\begin{equation}
	\left\llbracket Q \right\rrbracket = Q^+ - Q^- \, ,
\end{equation}
where the $Q^\pm$ are the values taken by the quantity $Q$ immediately on either side of the moving discontinuity.  In what follows, equations involving jumps should be understood to hold at the discontinuity and not elsewhere.

The spatial position of the jump must coincide for the two sides of the body, hence
\begin{equation}
	\left\llbracket \bX(s_0(t), t) \right\rrbracket = 0 \, .
\end{equation}
A total time derivative leads to a compatibility condition for velocities,
\begin{equation}
	\left\llbracket \partial_t\bX + \partial_ts_0\partial_s\bX \right\rrbracket = 0 \, , \label{compatibility}
\end{equation}
which could also be written with $\partial_ts_0$ moved outside the brackets\footnote{\label{notationfootnote}When applied to the material position vector $\bX(s,t)$, the symbols $\partial_s$ and $\partial_t$ respectively denote a partial derivative with respect to $s$ and a material time derivative.  When applied to the non-material point $s_0(t)$, $\partial_t$ is just a partial time derivative.}.  The quantity inside the brackets is simply the spatial velocity of the non-material point associated with the discontinuity.  If the boundary has continuous tangents, or is fixed ($\partial_ts_0 = 0$) in material coordinates, the condition \eqref{compatibility} implies continuity of the spatial velocity for the associated material point, $\left\llbracket \partial_t\bX \right\rrbracket = 0$.

In considering only inextensible strings of constant, uniform mass density, conservation of mass becomes relatively trivial.  The arc length coordinate is also a material coordinate, and the discontinuity exits one side of the body at the same rate it enters the other.  The motions of the body in the bulk on either side of the discontinuity, and the transfer of the line element across the discontinuity, must be isometries.  In the bulk this means that $\partial_t\left(\partial_s\bX\cdot\partial_s\bX\right)=0$, which is always true as $\partial_s\bX$ is a unit vector
.  Likewise, $\left.\left\llbracket \partial_s\bX\cdot\partial_s\bX \right\rrbracket\right. = 0$ at the jump.  Mass conservation will be examined in more detail in the following section.

Balances of momentum and energy may be obtained from a variational principle.  Consider an action of the form
\begin{align}
	A &= \int dt \,L = \int \!\!dt \int^{s_0(t)} \!\!ds \,\mathcal{L} \, ,\\
	2\mathcal{L} &= \mu\partial_t\bX\cdot\partial_t\bX - \sigma\left(\partial_s\bX\cdot\partial_s\bX - 1\right) \, , \label{stringaction}
\end{align}
representing the string on one side of the discontinuity.  The three unspecified limits of integration are irrelevant to the discussion.  The Lagrangian $L$ is that of an extended body with time-dependent boundary $s_0(t)$.  The Lagrangian density $\mathcal{L}$ is that of an inertial body of uniform mass density $\mu$, with $\sigma(s,t)$ a local multiplier enforcing inextensibility.  Upon variation of $\bX$, the bulk Euler-Lagrange equations of this density are the classical string equations.

The linear momentum equations are obtained from stationarity of $A$ under the spatial variation $\bX \rightarrow \bX + \delta\bX$.  Setting
\begin{equation}
	\delta A = \int \!\!dt \int^{s_0(t)} \!\!ds \, \frac{\delta\mathcal{L}}{\delta\bX}\cdot \delta\bX = 0 
\end{equation}
and using the Leibniz rule to manipulate time derivatives resulting from integration by parts, we obtain field equations for the bulk and the single relevant boundary:
\begin{align}
	&\mu\partial_t^2\bX - \partial_s\left(\sigma\partial_s\bX\right) = 0 \, , \label{bulkmomentum} \\
	&\left. \sigma\partial_s\bX + \mu\partial_ts_0\partial_t\bX = 0 \, \right|_{s=s_0} \, . \label{boundarymomentum}
\end{align}
For a boundary fixed with respect to material coordinates, $\partial_ts_0 = 0$ and we recover the usual material boundary condition for a string.  Noting that placing the time dependent point $s_0(t)$ at the lower limit of integration merely changes the sign of both boundary terms, we are led to a linear momentum jump condition for a moving, non-material, internal boundary,
\begin{equation}
	\left\llbracket \sigma\partial_s\bX + \mu\partial_ts_0\partial_t\bX \right\rrbracket = \bP \, , \label{momentumjump}
\end{equation}
where $\bP$ represents any supply of stress at the discontinuity \cite{OReillyVaradi99}, such as might be provided by an external obstacle.

There are no additional conditions for angular momentum.  A variation of the form $\bX \rightarrow \bX + \left(\bX - \bO\right) \times \delta\bC$, $\bO$ and $\delta\bC$ constant vectors, yields only redundant equations, as there is no jump in $\bX$ itself.

The energy equations are obtained from stationarity of $A$ under the temporal variation $t \rightarrow t+ \delta t$.  Although $\delta t$ is to be interpreted as a uniform time shift of the system, rather than a function of $s$, this fact is not to be invoked until near the end of the variational process, the point of which is to extract the temporal Noether current. The Lagrangian density \eqref{stringaction} is not an explicit function of time, but the Lagrangian is.  Setting
\begin{equation}
	\delta A = \int \!\!dt \, \frac{\delta L}{\delta t} \delta t = \int \!\!dt \left[ \left. \mathcal{L}\partial_ts_0\delta t  \right|_{s=s_0} + \int^{s_0(t)} \!\!ds \, \frac{\delta\mathcal{L}}{\delta\bX}\cdot \partial_t\bX\delta t \right]  = 0 
\end{equation}
and proceeding in a similar manner as before, we obtain bulk and boundary equations:
\begin{align}
	&\left[\mu\partial_t^2\bX - \partial_s\left(\sigma\partial_s\bX\right)\right]\cdot \partial_t\bX = 0 \, , \label{bulkenergy} \\
	&\left. \left[\sigma\partial_s\bX + \tfrac{1}{2}\mu\partial_ts_0\partial_t\bX\right]\cdot\partial_t\bX = 0 \, \right|_{s=s_0} \label{boundaryenergy} \, ,
\end{align}
where we have used the fact that $\partial_s\bX\cdot\partial_s\bX - 1 = 0$.  As this term is a constraint, there is no energy associated with it.  While the bulk equation \eqref{bulkenergy} and the fixed-boundary limit of the boundary equation \eqref{boundaryenergy} are redundant with respect to the momentum equations (\ref{bulkmomentum}-\ref{boundarymomentum}), the general boundary equation is not.  It leads to an energy jump condition,
\begin{equation}
	\left\llbracket \sigma\partial_s\bX\cdot\partial_t\bX + \tfrac{1}{2}\mu\partial_ts_0\partial_t\bX\cdot\partial_t\bX \right\rrbracket = E + \tfrac{1}{2}\bP\cdot\left(\partial_t\bX^+ + \partial_t\bX^-\right) \, , \label{energyjump}
\end{equation}
where $E$ represents any supply of power per area, including internal dissipation\footnote{The area unit is due to the cross section of the body, which is really the vanishing-radius limit of a three-dimensional string.}.  The form of the right hand side of \eqref{energyjump}, which makes use of the mean material velocity at the kink, is chosen so that the quantity $E$ is invariant under Galilean shifts $\bX \to \bX + \bC t$.  This is not a necessity.  One can imagine stress and power supplied by collision with a space-fixed obstacle external to the variational treatment of the body.  In such a treatment it might be simpler to write $E$ for the right hand side of condition \eqref{energyjump}\footnote{Note that Virga's comments \cite{Virga15} on the form of equation \eqref{energyjump} pertain to an earlier draft of this paper, when the right hand side was indeed simply written as $E$.}.  This would break Galilean invariance of the body, of $E$, and of the equations, even though the body-obstacle system must be invariant as a whole.  A general discussion of invariance restrictions on the supplies $\bP$ and $E$ may be found in O'Reilly and Varadi \cite{OReillyVaradi99}.

The invariant form of the jump conditions \eqref{momentumjump} and \eqref{energyjump} is that which would be derived from an action that explicitly includes boundary supply terms.  I have up to this point discussed only a one-sided bulk action, for simplicity of exposition.  The total Lagrangian leading to conditions \eqref{momentumjump} and \eqref{energyjump} has the form:
\begin{align}
	L &= \left. \mathcal{L}_S  \right|_{s=s_0} + \int_{s_0(t)} \!\!ds \,\mathcal{L}_V + \int^{s_0(t)} \!\!ds \,\mathcal{L}_V \, , \label{totalL}\\
	2\mathcal{L}_S &= \bP \cdot \left(\bX^+ + \bX^-\right) + 2Et \, , \\
	2\mathcal{L}_V &= \mu\partial_t\bX\cdot\partial_t\bX - \sigma\left(\partial_s\bX\cdot\partial_s\bX - 1\right) \, ,
\end{align}
where it is to be understood that $\bP$ and $E$ are unaffected by the variations.  Here the boundary density $\mathcal{L}_S$ is an explicit function of time, and the temporal variation must take this into account.  Note that the spatial velocity of the non-material discontinuity does not appear in the Lagrangian \eqref{totalL} or the balance \eqref{energyjump}, in contrast to Virga's configurational approach to energy balance \cite{Virga15}, which behaves identically under Galilean shifts.

\subsection{Comments}

The independence of momentum and energy supplies at a non-material boundary is reflected in the requirement of additional constitutive information to specify the behavior of dissipative shocks \cite{Virga15}, impacts \cite{HammGeminard10, Grewal11}, and phase transformations or other thermomechanical processes at moving discontinuities \cite{AbeyaratneKnowles90, PurohitBhattacharya02}.

Because the $\partial_t s_0$ term may be moved outside the brackets, the balances \eqref{momentumjump} and \eqref{energyjump} and velocity compatibility condition \eqref{compatibility} may be thought of as evolution equations for the boundary coordinate $s_0(t)$, a type of dynamics for a non-material point.  This viewpoint is sometimes adopted in studies of dynamic fracture, where an analogy with a peeling string was explored by Freund \cite{Freund90}.  Accelerations can be included in these dynamics as well, by considering an acceleration compatibility condition.

\section{A line discontinuity in a sheet}\label{sheet}

The results of this section are new, but they and the method of derivation are related to those of Hure and Audoly \cite{HureAudoly13} on the adhesion boundary of a static sheet on a surface.  I consider the case of a line discontinuity in a surface embedded in $\mathbb{E}^3$.  A line discontinuity in a two-dimensional body in $\mathbb{E}^2$ is a surface itself (in the sense of being codimension-1 with respect to $\mathbb{E}^2$), and may be addressed in the same manner as surface discontinuities in three-dimensional bodies in $\mathbb{E}^3$.  The latter is a classical topic, a representative example of which may be found in chapters 32-33 of the recent text by Gurtin, Fried, and Anand \cite{GurtinFriedAnand10}.  Pop and Wang \cite{PopWang81} considered higher-order line discontinuities, with continuous tangent vectors, propagating in two-dimensional bodies in $\mathbb{E}^3$.  They used convected coordinates, and assumed that one side of the surface was quiescent.

A line discontinuity may be closed, terminate on the boundary of the surface, or terminate at point discontinuities within the surface.  Let us neglect the third option for the moment.  Figure \ref{surface} depicts the second option, with discontinuities in some of the first derivatives of the position vector.  The body $\bX\left(\{\lambda^\alpha\}, t\right)$ is now parameterized by a set of global material coordinates $\{\lambda^\alpha\}$, $\alpha \in \{1,2\}$, which span the discontinuity, and by the time $t$.  Use of such a coordinate system facilitates description, and does not restrict us in considering the type of deformations relevant to inextensible thin bodies.  The discontinuity is now a curve with surface coordinates $\{\lambda^\alpha_0 (l, t)\}$ parameterized by a curve coordinate $l$ and the time $t$.  For simplicity, let us assume that $l$ is independent of time, rather than something like an arc length.  The curve derivative is $\partial_l = \partial_l\lambda^\alpha_0\partial_\alpha\,$.  

First order discontinuities are special in that, unless they are restricted, they will result in a jump in the surface metric tensor $a_{\alpha\beta} \equiv \partial_\alpha\bX\cdot\partial_\beta\bX$.  However, certain restrictions are reasonable when considering inextensible motions of sheets.  I seek to model such discontinuities as may, for example, appear dynamically during the motion of an initially smooth surface.  Let us consider ``$\mathcal{C}^1$-foldable'' discontinuities, defined such that the discontinuous surface is isometric to some $\mathcal{C}^1$-smooth configuration\footnote{The exception to this statement is on the discontinuity itself, where the metric is not defined.  I am requiring that the surfaces on either side can be bent, but not stretched, in such a manner that the discontinuity in the tangents disappears.}.  On this latter configuration, I paint an initial set of $\mathcal{C}^1$-smooth global coordinate curves, and admit a velocity field tangential to the surface that advects these coordinates while preserving the metric (a Killing flow).  These coordinates and surface velocities will be carried over under the isometry to the discontinuous surface, and intrinsically defined surface quantities will not suffer a jump. For example, the angle within the surface between a coordinate curve and the discontinuity curve remains the same on either side.  It should be mentioned that the foldability condition puts additional strict constraints on the relationship between extrinsic properties of surfaces on either side of a fold \cite{DuncanDuncan82}.  The difficult but important tasks of constructing an appropriate surface pair and finding a Killing flow are beyond the scope of this paper.

It will be helpful to envision the orthonormal Darboux frames defined by the discontinuity curve and the two surfaces on either side of it.  At each $l$, such a frame consists of a (curve and surface) tangent $\uvc{\theta} = \frac{\partial_l\bX}{\| \partial_l\bX \|}$, a (surface) normal $\uvc{N}^\pm$, and a (surface) tangent (curve) normal $\uvc{\nu}^\pm = \uvc{\theta} \times \uvc{N}^\pm$.  The tangent is shared by both sides, 
while the other vectors may differ across the discontinuity, as indicated by the notation.  It is the tangent normal $\uvc{\nu}^\pm$ that will be physically important, and the fact that it has a jump leads to significant differences between the present system and discontinuities in space-filling bodies.

Quantities on either side of the jump are defined as follows,
\begin{equation}
	Q^\pm(\{\lambda^\alpha_0(l,t)\},t) \equiv \lim_{\epsilon\to 0} Q\left( \{\lambda^\alpha_0(l,t) \pm \epsilon \nu^{\alpha}(l,t)\}, t \right) \, , 
\end{equation}
where the components $\nu^\alpha$ of the tangent normal are identical on either side due to the $\mathcal{C}^1$-foldability condition.

Not only the position $\bX$, but its $l$-derivatives, must coincide on either side of the jump, hence
\begin{align}
	\left\llbracket \bX\left(\{\lambda^\alpha_0(l,t)\}, t\right) \right\rrbracket &= 0 \, , \label{xjumpsurf} \\
	\left\llbracket \partial_l \bX \right\rrbracket &= 0 \, , \label{dlxjumpsurf} \\
	\left\llbracket \partial^2_l \bX \right\rrbracket &= 0 \, , \label{d2lxjumpsurf}
\end{align}
and so on, with \eqref{dlxjumpsurf} also implying continuity of the curve metric
\begin{equation}
		\left\llbracket \partial_l\bX\cdot\partial_l \bX \right\rrbracket = 0 \, . \label{curvemetricjumpsurf}
\end{equation}
Again, a total time derivative of \eqref{xjumpsurf} leads to a compatibility condition relating the spatial velocity of the body to the 
surface velocity of the discontinuity within the body coordinates,
\begin{equation}
	\left\llbracket \partial_t\bX + \partial_t\lambda^\alpha_0\partial_\alpha\bX \right\rrbracket = 0 \, . \label{velocityjumpsurf}
\end{equation}
However, for a $\mathcal{C}^1$-foldable discontinuity, the second term cannot have a jump in its projection onto the surface tangents.  Hence,
\begin{align}
	\left\llbracket  \partial_t\lambda^\alpha_0\partial_\alpha\bX\cdot\partial_l\bX \right\rrbracket &= 0 \, , \label{velocityprojtanjumpsurf} \\
	\left\llbracket  \partial_t\lambda^\alpha_0\partial_\alpha\bX\cdot\uvc{\nu} \right\rrbracket &= 0 \, .
\label{velocityprojnormjumpsurf} \end{align}
The bracketed quantity in the second expression \eqref{velocityprojnormjumpsurf} is just the velocity of the boundary along the tangent normal.  
The first expression \eqref{velocityprojtanjumpsurf} can be used, along with matching of the curve tangent, to modify the second term in \eqref{velocityjumpsurf} so that
\begin{equation}
	\left\llbracket \partial_t\bX + \partial_t\lambda^\alpha_0\partial_\alpha\bX\cdot\uvc{\nu}\uvc{\nu}\right\rrbracket = 0 \, . \label{velocityjumpsurf2}
\end{equation}
The bracketed quantity in \eqref{velocityjumpsurf2} is the spatial velocity of the non-material discontinuity.  The conditions \eqref{dlxjumpsurf} and \eqref{velocityjumpsurf2} are compatibility \emph{\`{a} la Hadamard} \cite{GurtinFriedAnand10}, a standard requirement in continuum mechanics.  Note, however, an important difference between the present system and a codimension-1 discontinuity in a space-filling body.  The relevant normal vector is the tangent normal, which is the surface, rather than the spatial, complement of the tangent space of the discontinuity.  We cannot move the tangent normal $\uvc{\nu}$ outside the brackets, as it suffers a jump; pointing out of one side of the body does not imply pointing into the other side.  In the formalism of Gurtin, Fried, and Anand \cite{GurtinFriedAnand10}, there is a jump in the projection operator.  On the other hand, we may move $\partial_t\lambda^\alpha_0$ and $\partial_\alpha\bX\cdot\uvc{\nu} = \nu_\alpha$ outside the brackets, because of the coordinate system we have chosen on the surface.  In this sense, the situation described here is less general than one which allows for discontinuities in coordinate lines in the surface direction normal to the discontinuity, as is appropriate for some types of material deformations \cite{GurtinFriedAnand10}.

The isometry preserves geodesic curvature, and its equivalents for other coordinates like $l$, hence
\begin{equation}
	\left\llbracket \partial^2_l\bX\cdot\uvc{\nu} \right\rrbracket = 0 \, . \label{geodesicjumpsurf} \\
\end{equation}
Conditions \eqref{curvemetricjumpsurf} and \eqref{geodesicjumpsurf} are compatibility \emph{\`{a} la Darmois} \cite{BonnorVickers81}, a common imposition on codimension-1 discontinuities in general relativity.  Again the present system, with its second embedding space, is distinctive in the substitution of the tangent normal for the spatial curve normal.  Note that the Darmois conditions, unlike those of Hadamard, could be described in terms purely intrinsic to the surface.

Consideration of an inextensible sheet with constant, uniform mass density makes the treatment of mass conservation considerably easier than it could be.  The $\mathcal{C}^1$-foldability condition ensures that $\left\llbracket a_{\alpha\beta} \right\rrbracket = 0$, which along with the uniformity of $\mu$ implies both $\left\llbracket \mu a_{\alpha\beta} \right\rrbracket = 0$ and $\left\llbracket \,\mu \sqrt{a}\, \right\rrbracket = 0$.  The latter is equivalent to the expression of mass conservation in O'Reilly and Varadi \cite{OReillyVaradi99}.  For this statement to be interpreted as covariant, $\mu$ must transform nontensorially as a scalar  \emph{density}, reciprocally to the square root of the metric determinant $\sqrt{a}$.

It may be helpful to look at the system from the perspective of the usual continuum mechanical derivation of mass conservation using a transport relation for a material volume. Consider the total time derivative of an integral of a quantity $Q$ taken over a (two-dimensional) material volume $V(t)$ of a surface with time-independent metric $a_{\alpha\beta}$ and the discontinuity as its boundary $\partial V(t)$,
\begin{equation}
	\frac{d}{dt}\int_{V(t)}\!\!dV \, Q  = \int_{V(t)}\!\!dV \, \partial_t Q + \int_{\partial V(t)}\!\!dS \, Q \partial_t\lambda^\alpha_0 \nu_\alpha \, , \label{transport}
\end{equation}
where $dV = \sqrt{a}\prod\limits_{\alpha} d\lambda^\alpha$ and $dS = \| \partial_l \bX \| dl$.  If $Q$ is a uniform mass density $\mu$, then \eqref{dlxjumpsurf} and \eqref{velocityprojnormjumpsurf} ensure that it will be conserved when matching the boundary term with that of the complementary part of the surface, for which the relevant normal has the opposite sense.

In considering systems undergoing elastic or plastic deformations across the discontinuity, one must allow for time-dependence and jumps of the body metric, and take greater care in treating mass conservation.  A general treatment of surfaces with time-dependent metrics requires not only covariance of the jump condition, but also of the total time derivative \cite{Thiffeault01} of any quantity that lives at least partially in the surface, such as the metric tensor, surface coordinates, or surface derivatives of the position vector.  

Let us obtain balances of momentum and energy from an action of the form
\begin{align}
	A &= \int dt \,L = \int \!\!dt \int_{V(t)} \!\!dV \,\mathcal{L} \, ,\\
	2\mathcal{L} &= \mu\partial_t\bX\cdot\partial_t\bX - \sigma^{\alpha\beta}\left(\partial_\alpha\bX\cdot\partial_\beta\bX - a_{\alpha\beta}\right) \, , \label{sheetaction}
\end{align}
where the local multiplier $\sigma^{\alpha\beta}$ preserving the surface metric is a surface stress tensor.  The form of \eqref{sheetaction}, of which the string form \eqref{stringaction} is a special case, does not depend on dimensionality of the body.

We proceed as before. The linear momentum equations are obtained from stationarity of $A$ under the spatial variation $\bX \rightarrow \bX + \delta\bX$.  Setting
\begin{equation}
	\delta A = \int \!\!dt \int_{V(t)} \!\!dV \, \frac{\delta\mathcal{L}}{\delta\bX}\cdot \delta\bX = 0 
\end{equation}
and using \eqref{transport} to manipulate time derivatives, we obtain for the bulk and boundary:
\begin{align}
	&\mu\partial_t^2\bX - \nabla_\alpha\left(\sigma^{\alpha\beta}\partial_\beta\bX\right) = 0 \, , \label{bulkmomentumsurf} \\
	&\left. \nu_\alpha\left[\sigma^{\alpha\beta}\partial_\beta\bX + \mu\partial_t\lambda^\alpha_0 \partial_t\bX\right] = 0 \, \right|_{\partial V} \, , \label{boundarymomentumsurf}
\end{align}
where $\nabla$ is a covariant derivative on the surface.  The easiest way to obtain these equations is to exploit the fact that $\delta\bX$, like $\bX$, has no surface indices and behaves like a surface scalar: $\partial_\alpha\delta\bX = \nabla_\alpha\delta\bX$.  Note that if the boundary $\partial V$ terminates in a point or corner, it itself has a boundary $\partial\partial V$, though this is not relevant to the action considered here.  For a fixed-coordinate boundary, we recover the usual material boundary condition for a sheet.  Matching with the complementary part of the surface provides a linear momentum jump condition, 
\begin{equation}
	\left\llbracket \nu_\alpha\sigma^{\alpha\beta}\partial_\beta\bX + \mu\partial_t\lambda^\alpha_0 \nu_\alpha \partial_t\bX \right\rrbracket = \bP \, . \label{momentumjumpsurf}
\end{equation}

There are no additional conditions for angular momentum.

The energy equations are obtained from stationarity of $A$ under the temporal variation $t \rightarrow t+ \delta t$.  Again $\mathcal{L}$ is not explicitly time-dependent.  Setting
\begin{equation}
	\delta A = \int \!\!dt \, \frac{\delta L}{\delta t} \delta t = \int \!\!dt \left[ \int_{\partial V(t)} \!\!dS \, \mathcal{L}\partial_t\lambda^\alpha_0\nu_\alpha\delta t  + \int_{V(t)} \!\!dV \frac{\delta\mathcal{L}}{\delta\bX}\cdot \partial_t\bX\delta t \right]  = 0 
\end{equation}
and remembering that $\partial_\alpha\bX\cdot\partial_\beta\bX - a_{\alpha\beta} = 0$, we obtain bulk and boundary equations:
\begin{align}
	&\left[\mu\partial_t^2\bX - \nabla_\alpha\left(\sigma^{\alpha\beta}\partial_\beta\bX\right)\right]\cdot\partial_t\bX = 0 \, , \label{bulkenergysurf} \\
	&\left. \nu_\alpha\left[\sigma^{\alpha\beta}\partial_\beta\bX + \tfrac{1}{2}\mu\partial_t\lambda^\alpha_0 \partial_t\bX\right]\cdot\partial_t\bX = 0 \, \right|_{\partial V} \, . \label{boundaryenergysurf}
\end{align}
The bulk equation \eqref{bulkenergysurf} and the fixed-boundary limit of the boundary equation \eqref{boundaryenergysurf} are redundant with respect to the momentum equations (\ref{bulkmomentumsurf}-\ref{boundarymomentumsurf}).  The general boundary equation leads to an energy jump condition,
\begin{equation}
	\left\llbracket \nu_\alpha\sigma^{\alpha\beta}\partial_\beta\bX\cdot\partial_t\bX + \tfrac{1}{2}\mu\partial_t\lambda^\alpha_0 \nu_\alpha\partial_t\bX\cdot\partial_t\bX \right\rrbracket = E + \tfrac{1}{2}\bP\cdot\left(\partial_t\bX^+ + \partial_t\bX^-\right) \, , \label{energyjumpsurf}
\end{equation}
where again the right hand side is chosen so as to preserve invariance under Galilean shifts.

\subsection{Comments}

At this point, one might ask whether generic moving first order discontinuities, even of the restricted type considered thus far, are consistent with inextensible motions of sheets.  The $\mathcal{C}^1$-foldable construction is common when considering static curved folds \cite{DuncanDuncan82, DiasSantangelo12}, but in such cases the understanding is that the fold carries some kind of singularity in the metric.  Gauss's \emph{theorema egregium}, though not strictly applicable, seems to preclude general discontinuities in tangent vectors without stretching of the surface, and presumably the latter should be associated with some energy and surface stresses.  This issue will also arise for intersections of line defects, line defects with discontinuities in the curve tangents, or line defects terminating at point defects inside the sheet.  If the stretching is plastic, defects will likely be pinned, as one can see in manipulating a paper sheet.  If elastic, they will likely be mobile, and the energy contained in the jump could manifest as an effective line tension acting to shorten or resist lengthening of the discontinuity.  One could imagine that the effects should appear not just in the power per area supply $E$, but in the stress supply $\bP$ as a traction proportional to the geodesic curvature vector.  These several difficulties could in principle be avoided by considering an in-surface elastic term in the action instead of a constraint on the metric, so as to relax the infinite energy penalty for stretching the sheet.  However, this would not only add a new term to the energy jump condition, but complicate the analysis considerably by causing the metric to be time-dependent and thus requiring covariant treatment of time derivatives.  Isometric approximations work extremely well for static problems involving curved discontinuities or other defects with localized stretching \cite{DuncanDuncan82, Witten07, DiasSantangelo12, Guven13dipoles}, and for quasistatic models of crack propagation in brittle thin sheets \cite{Roman13}.  Whether or not this remains true in dynamic regimes remains to be determined.

\section{Point discontinuities in line discontinuities in a sheet, and a crack as a time-dependent boundary}\label{crack}

Line defects may terminate in point defects inside a sheet, or contain internal kinks of their own.  Crack tips in sheets are point defects constituting unique non-material points on otherwise material external boundaries.  To treat either type of point defect, we must consider boundaries $\partial V$ that have boundaries $\partial\partial V$, and the corresponding boundary-boundary velocities $\partial_t\lambda^\alpha_0 \theta_\alpha$ along the tangents 
normal to $\partial\partial V$.  

Let us consider a general Lagrangian with both boundary and bulk terms,
\begin{equation}
	L = \int_{\partial V(t)} \!\!dS \,\mathcal{L}_S + \int_{V(t)} \!\!dV \,\mathcal{L}_V \, . \label{generalL}
\end{equation}
The actions we have been considering until now are bulk actions that depend at most on first derivatives of $\bX$, meaning that no ``boundary-boundary'' terms arise to provide forces on the ends of discontinuity curves.  This would no longer be the case, for example, for a finite-thickness sheet action that includes bulk bending energy terms, which would be higher order in surface derivatives.  Boundary-boundary velocity terms might contribute to such end forces if the bulk contribution to the action contains mixed time and surface derivatives, 
or if there is a boundary contribution to the action.

Relations such as \eqref{transport} hold for piecewise smooth boundaries.  If the densities $\mathcal{L}_S$ and $\mathcal{L}_V$ are not explicit functions of time, the spatial and temporal variations of the Lagrangian \eqref{generalL} take the forms
\begin{align}
	&\frac{\delta L}{\delta \bX} \cdot \delta \bX =  \int_{\partial V(t)} \!\!dS \, \frac{\delta\mathcal{L}_S}{\delta\bX}\cdot \delta\bX + \int_{V(t)} \!\!dV \, \frac{\delta\mathcal{L}_V}{\delta\bX}\cdot \delta\bX \, , \label{spatiallagrangian} \\ 
	&\frac{\delta L}{\delta t} \delta t =  \left. \mathcal{L}_S\partial_t\lambda^\alpha_0\theta_\alpha\delta t  \right|_{\partial\partial V(t)}  + \int_{\partial V(t)} \!\!dS \left(\frac{\delta\mathcal{L}_S}{\delta\bX}\cdot \partial_t\bX\delta t + \mathcal{L}_V\partial_t\lambda^\alpha_0\nu_\alpha\delta t \right) + \int_{V(t)} \!\!dV \,\frac{\delta\mathcal{L}_V}{\delta\bX}\cdot \partial_t\bX\delta t \, . \label{temporallagrangian}
\end{align}
Terms may move from bulk $V$ to boundary $\partial V$ to boundary-boundary $\partial\partial V$ if they are divergences; the appropriate divergence on $\partial V$ is an $l$-derivative.  
Additionally, source terms could exist on $\partial\partial V$, for example if a slicing tool is responsible for crack propagation, or if energy is dissipated at the point discontinuity.

For an internal point defect on the end of an internal line defect, the boundary $\partial\partial V$ of the ``boundary'' $\partial V$ is simply a point with a single unit tangent normal $\uvc{\theta}$.  The boundary density $\mathcal{L}_S$ might represent an energy associated with localized stretching along a fold.

\begin{figure}[here]
	\begin{overpic}[width=2in]{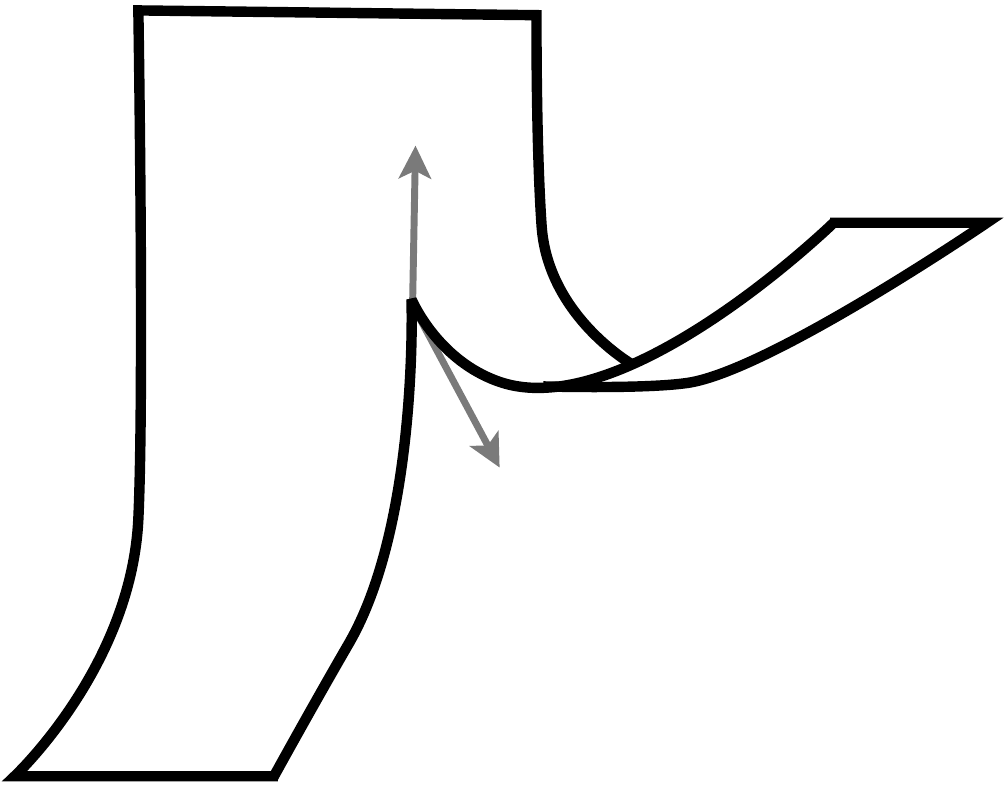}
	\put(50,90){$\uvc{\theta}^L$}
	\put(70,35){$\uvc{\theta}^R$}
	\put(25,70){$\partial\partial V(t)$}
	\end{overpic}
\caption{Crack geometry in a sheet, with boundary tangents that are normal to a boundary-boundary point.  Details and definitions in text.}
\label{cracktear}
\end{figure}

The situation is a bit more complicated for a crack tip (Figure \ref{cracktear}), or a kink in an internal boundary.  The tip $\partial\partial V$ is the intersection, at some non-material boundary point $l_0 (t)$, of two pieces of boundary $\partial V$.  Each piece carries its own tangent normal $\uvc{\theta}^\RL$, which is the corresponding boundary tangent immediately on either side of $l_0 (t)$.  While the tip $\partial\partial V$ is a single point with single-valued surface velocity $\partial_t\lambda^\alpha_0$, there are two values of $\partial_\alpha\bX\cdot\uvc{\theta}^\RL = \theta_\alpha^\RL$.  The first term of the temporal variation of $L$ will take the form $\mathcal{L}_S\partial_t\lambda^\alpha_0\left(\theta_\alpha^R - \theta_\alpha^L\right)$. 
For a crack, $\partial V$ is a free material boundary, and $\mathcal{L}_S$ is the cost of creating more $\partial V$.  While this boundary is time-dependent, the bulk $V$ is not, and $\partial_t\lambda^\alpha_0\nu_\alpha = 0$.  
Another feature of a crack is that $\theta_\alpha^R = -\theta_\alpha^L$.  

Crack propagation in thin sheets is often geometrically and dynamically interesting \cite{Roman13, VandenbergheVillermaux13}.  One example is ``trouser'' tearing, a Mode III fracture test in which two ``legs'' are pulled in opposite directions perpendicular to the remainder of a flat sheet.  In considering such a geometry for sheets without bending resistance, it may be necessary to imagine the crack tip sitting at the nexus of two material boundaries along the ``inseams'' and two internal line defects parallel to and just below the ``waist''.


%

\section{Examples of sheet motions near a line discontinuity}\label{examples}

I present two general examples for sheet motions with a line discontinuity, which are easily applicable to strings as a special case.  Let there be two moving surfaces which coincide on the discontinuity.  The particular forms of the surfaces may remain unspecified, as long as they admit the surface velocities of the solutions as boundary values of Killing fields. Let coordinates be chosen such that surface velocities and discontinuity tangent normals are parallel to one set of coordinate lines, with index $n$. This means that the discontinuity must be along a geodesic, and that flow of material is strictly normal to it.  Recall that there is no jump in the surface metric across the discontinuity, hence $\left\llbracket \partial_n\bX\cdot\partial_n\bX \right\rrbracket = 0$.  Let there be no external forces or internal dissipation, so that $\bP = 0$, $E = 0$.

 It may be useful to think of a fold propagating through a chain, rug, or bedsheet.  Either part of Figure \ref{jumps}, without coordinate lines, may serve as illustration.  I am, however, only considering a portion of the surface adjacent to the discontinuity, and whether such solutions may be extended indefinitely into the bulk on either side depends on the problem at hand.  Any bulk solutions stitched to the jump solutions below must satisfy the bulk field equations, and possibly additional boundary conditions some distance away.
  
\subsection{Steady configuration}

Let the surfaces, or those portions near the discontinuity, move tangentially to themselves in the $n$ direction with velocity $T$, so that material flows along a fixed shape in an inertial frame.  The solution
\begin{align}
	\partial_t\bX^\pm &= T\partial_n\bX^\pm \, , \\
	\partial_t\lambda^n_0 &= -T \, , \\
	\sigma^{nn\pm} &= \mu T^2 \, ,
\end{align}
satisfies \eqref{velocityjumpsurf}, \eqref{momentumjumpsurf}, and \eqref{energyjumpsurf}, with terms vanishing identically for the latter equation.  Thus the Routh ``lariat'' solution, applicable to continuous strings \cite{Routh55, HealeyPapadopoulos90} and some continuous surfaces \cite{Guven13skirts}, obtains also in the discontinuous limit.  The discontinuity, like the rest of the non-material shape, does not move.

\subsection{Semi-quiescent configuration}

Let one surface, or its portion near the discontinuity, be static.  The solution
\begin{align}
	\partial_t\bX^+ &= 0 \, , \\
	\partial_t\bX^- &= -U\left\llbracket\partial_n\bX\right\rrbracket \, , \\
	\partial_t\lambda^n_0 &= -U \, , \\
	\sigma^{nn\pm} &= \mu U^2 \, ,
\end{align}
satisfies \eqref{velocityjumpsurf}, \eqref{momentumjumpsurf}, and \eqref{energyjumpsurf}, and includes the falling folded vertical chain \cite{Schagerl97} as a special case.  This is also the relevant geometry for peeling adhesive tape.  Note that $\left\llbracket\partial_n\bX\right\rrbracket$ always points ``in between'' the two surfaces on the concave side.  The motion of the discontinuity in the laboratory frame is $-U\partial_n\bX^+$, and the quiescent surface simply extends or recedes along its tangent.  This result may seem counterintuitive; momentum perpendicular to the quiescent piece seems to disappear or appear from nowhere.  However, momentum is conserved, just as it is for the discontinuous lariat, where the stress resultant redirects elements of the body as they pass through the kink.  Here the elements are brought to rest or into motion in the same manner.

\subsection{Comments}

These solutions, like the continuous Routh lariat, are surface velocity- and shape-agnostic in that they do not specify anything about the velocities $T$ and $U$ or configuration $\bX$.  However, for nontrivial $T$ and $U$ they cannot be extended in time until the discontinuity exits a finite-sized body, as the free material boundary condition will be incompatible with nonzero $\sigma$ at the jump.  The behavior of a flexible body near its free boundary is not currently understood even in the absence of discontinuities, so the latter cannot be blamed for the difficulty.

\section{A point discontinuity in an \emph{elastica}}\label{elastica}

The Euler \emph{elastica} is a common model for thin elastic bodies in effectively planar situations. 
The body is symmetric about its midline $\bX$, which carries a single director field that may be explicitly related to derivatives of $\bX$.  An appropriate Lagrangian density, written purely in terms of $\bX$, is
\begin{equation}
	2\mathcal{L} = \mu\partial_t\bX\cdot\partial_t\bX + M\partial_t\partial_s\bX\cdot\partial_t\partial_s\bX - B\partial_s^2\bX\cdot\partial_s^2\bX - \sigma\left(\partial_s\bX\cdot\partial_s\bX - 1\right) \, , \label{elasticalagrangiandensity}
\end{equation}
where $B$ is a bending modulus and $M$ a moment of inertia density.  In the plane, there is no need to consider twist of the body, and the rate of rotation of the directors is simply that of the tangents $\partial_s\bX$.  Proceeding as for the string leads to the bulk field equation
\begin{equation}
	\mu\partial_t^2\bX - \partial_s\left(M\partial_t^2\partial_s\bX - B\partial_s^3\bX + \sigma\partial_s\bX \right) = 0 \, , \label{bulkmomentumelastica} 
\end{equation}
and jump conditions
\begin{align}
	\left\llbracket M\partial_t^2\partial_s\bX -B\partial_s^3\bX + \sigma\partial_s\bX +\mu\partial_ts_0\partial_t\bX \right\rrbracket &= \bP \, , \label{momentumjumpelastica} \\
	\left\llbracket B\partial_s^2\bX + M\partial_ts_0\partial_t\partial_s\bX \right\rrbracket &= \bR \, , \label{momentumjumpelasticatorque} \\
	\left\llbracket B\partial_s^2\bX\cdot\partial_t\partial_s\bX+ \left(M\partial_t^2\partial_s\bX -B\partial_s^3\bX + \sigma\partial_s\bX\right)\cdot\partial_t\bX \right.\quad\quad& \nonumber \\
	\left. + \tfrac{1}{2}\partial_ts_0\left(\mu\partial_t\bX\cdot\partial_t\bX + M\partial_t\partial_s\bX\cdot\partial_t\partial_s\bX + B\partial_s^2\bX\cdot\partial_s^2\bX\right) \right\rrbracket &= E + \tfrac{1}{2}\bP\cdot\left(\partial_t\bX^+ + \partial_t\bX^-\right)  \, , \label{energyjumpelastica} 
\end{align}
where $\bR$ is a supply of torque per area.  Because the action depends on second derivatives of position, there are jump conditions corresponding to $\delta\bX$, $\partial_s\delta\bX$, and $\delta t$.   To obtain these equations, we privilege the time derivative when integrating by parts, and as a last manipulation invoke the facts that $\partial_s\left(\partial_t\bX\delta t\right) = \partial_s\partial_t\bX\delta t = \partial_t\partial_s\bX\delta t$.  

These results are consistent with those of O'Reilly \cite{OReilly07} for general rod models, but the absence of explicit director dependence of the Lagrangian density \eqref{elasticalagrangiandensity} means that we have no director momentum equation. The correspondence with O'Reilly's paper \cite{OReilly07}, or the standard Cosserat formulation of the Kirchhoff equations for rods, tells us that the ``contact force'' for moving elastica is related not only to our tension $\sigma$ and squared curvature $\partial^2_s\bX\cdot\partial^2_s\bX$, but to an angular momentum term corresponding to the accelerations of the tangent vectors or, equivalently for an inextensible curve, their squared rate of change: $\partial_t\partial_s\bX\cdot\partial_t\partial_s\bX = - \partial_t^2\partial_s\bX\cdot\partial_s\bX$.  The material boundary terms in \eqref{energyjumpelastica} represent two contributions to the power.  One comes from the dot product of force and velocity, the other from torque and angular velocity as embodied in the rates of change of the tangent vectors $\partial_t\partial_s\bX$.  For non-moving, material, internal boundaries, the velocity compatibility condition \eqref{compatibility} implies continuity of the spatial velocity $\partial_t\bX$.  If, in addition, the angle at this internal boundary is fixed, so that there is continuity of $\partial_t\partial_s\bX$, the energy jump condition \eqref{energyjumpelastica} is no longer independent of \eqref{momentumjumpelastica} and \eqref{momentumjumpelasticatorque}.  Alternately, if the tangents $\partial_s\bX$ are continuous, the velocity compatibility condition \eqref{compatibility} again implies continuity of $\partial_t\bX$, and the conditions simplify to:
\begin{align}
	\left\llbracket M\partial_t^2\partial_s\bX -B\partial_s^3\bX \right\rrbracket + \left\llbracket \sigma\right\rrbracket \partial_s\bX &= \bP \, ,  \\
	\left\llbracket B\partial_s^2\bX + M\partial_ts_0\partial_t\partial_s\bX \right\rrbracket &= \bR \, ,  \\
	\left\llbracket B\partial_s^2\bX\cdot\partial_t\partial_s\bX \right\rrbracket +  \left\llbracket \tfrac{1}{2}\partial_ts_0\left( M\partial_t\partial_s\bX\cdot\partial_t\partial_s\bX + B\partial_s^2\bX\cdot\partial_s^2\bX \right) \right\rrbracket &= E \, , \label{energyjumpelasticasmooth}
\end{align}
where a term of the form $\bP\cdot\partial_t\bX$ has been cancelled from either side of \eqref{energyjumpelasticasmooth}.
This may be the appropriate set of conditions for many moving contact problems involving thin elastic bodies.

\section{Summary and comments}

I have presented a method for derivation of momentum and energy jump conditions from an action principle, and indicated avenues for generalization.  The process also leads quite naturally to a nonstandard approach to boundary conditions at a crack tip.  I focused primarily on discontinuities in inextensible strings and sheets, though conditions for the \emph{elastica} were also derived.  The results for sheets are new, and are embellished by two solutions for moving line defects.  

A useful extension of this work would be a generalization to time-dependent body metrics, to allow for intrinsic deformations of the body both outside and within the discontinuities.  Phenomena that could be treated in this manner include fracture with elastic energy release, propagating plastic instabilities, sheet metal drawing, and solid$\to$solid phase transformations.  Extension to dissipative systems subject to variational principles should be straightforward.  The present approach can presumably be formalized in a four-dimensional setting that treats time as a fourth coordinate and involves only a single variation.

Another interesting extension would be to reformulate the jump conditions for line discontinuities as evolution equations for a non-material curve, written in terms of geometric quantities associated with the curve and its two embedding surfaces.

\section*{Acknowledgments}

I thank O M O'Reilly for suggesting a variational treatment of singular sources which led to insight about Galilean shifts, C D Santangelo for an instrumental early discussion, and E G Virga who kindly shared notes on a dissipation principle and on a Cosserat approach to planar elastic curves.  I thank J Guven and T R Powers for informing me of some references.  I am also grateful for the hospitality of the FAST lab of the Universit\'{e} Paris-Sud in Orsay, where the first draft of this work was completed.

\bibliographystyle{unsrt}


\end{document}